\documentstyle[twoside,psfig]{article}

\catcode`\@=11
\long\def\@makefntext#1{
\protect\noindent \hbox to 3.2pt {\hskip-.9pt  
$^{{\eightrm\@thefnmark}}$\hfil}#1\hfill}               

\def\thefootnote{\fnsymbol{footnote}}
\def\@makefnmark{\hbox to 0pt{$^{\@thefnmark}$\hss}}    
        
\def\ps@myheadings{\let\@mkboth\@gobbletwo
\def\@oddhead{\hbox{}
\rightmark\hfil\eightrm\thepage}   
\def\@oddfoot{}\def\@evenhead{\eightrm\thepage\hfil
\leftmark\hbox{}}\def\@evenfoot{}
\def\sectionmark##1{}\def\subsectionmark##1{}}

\oddsidemargin=\evensidemargin
\renewcommand{\thefootnote}{\fnsymbol{footnote}}

\newcounter{sectionc}\newcounter{subsectionc}\newcounter{subsubsectionc}
\renewcommand{\section}[1] {\vspace{12pt}\addtocounter{sectionc}{1} 
\setcounter{subsectionc}{0}\setcounter{subsubsectionc}{0}\noindent 
        {\tenbf\thesectionc. #1}\par\vspace{5pt}}
\renewcommand{\subsection}[1] {\vspace{12pt}\addtocounter{subsectionc}{1} 
        \setcounter{subsubsectionc}{0}\noindent 
        {\bf\thesectionc.\thesubsectionc. {\kern1pt \bfit #1}}\par\vspace{5pt}}
\renewcommand{\subsubsection}[1] {\vspace{12pt}\addtocounter{subsubsectionc}{1}
        \noindent{\tenrm\thesectionc.\thesubsectionc.\thesubsubsectionc.
        {\kern1pt \tenit #1}}\par\vspace{5pt}}
\newcommand{\nonumsection}[1] {\vspace{12pt}\noindent{\tenbf #1}
        \par\vspace{5pt}}

\newcounter{appendixc}
\newcounter{subappendixc}[appendixc]
\newcounter{subsubappendixc}[subappendixc]
\renewcommand{\thesubappendixc}{\Alph{appendixc}.\arabic{subappendixc}}
\renewcommand{\thesubsubappendixc}
        {\Alph{appendixc}.\arabic{subappendixc}.\arabic{subsubappendixc}}

\renewcommand{\appendix}[1] {\vspace{12pt}
        \refstepcounter{appendixc}
        \setcounter{figure}{0}
        \setcounter{table}{0}
        \setcounter{lemma}{0}
        \setcounter{theorem}{0}
        \setcounter{corollary}{0}
        \setcounter{definition}{0}
        \setcounter{equation}{0}
        \renewcommand{\thefigure}{\Alph{appendixc}.\arabic{figure}}
        \renewcommand{\thetable}{\Alph{appendixc}.\arabic{table}}
        \renewcommand{\theappendixc}{\Alph{appendixc}}
        \renewcommand{\thelemma}{\Alph{appendixc}.\arabic{lemma}}
        \renewcommand{\thetheorem}{\Alph{appendixc}.\arabic{theorem}}
        \renewcommand{\thedefinition}{\Alph{appendixc}.\arabic{definition}}
        \renewcommand{\thecorollary}{\Alph{appendixc}.\arabic{corollary}}
        \renewcommand{\theequation}{\Alph{appendixc}.\arabic{equation}}
        \noindent{\tenbf Appendix \theappendixc #1}\par\vspace{5pt}}
\newcommand{\subappendix}[1] {\vspace{12pt}
        \refstepcounter{subappendixc}
        \noindent{\bf Appendix \thesubappendixc. {\kern1pt \bfit #1}}
        \par\vspace{5pt}}
\newcommand{\subsubappendix}[1] {\vspace{12pt}
        \refstepcounter{subsubappendixc}
        \noindent{\rm Appendix \thesubsubappendixc. {\kern1pt \tenit #1}}
        \par\vspace{5pt}}

\newcommand{\textlineskip}{\baselineskip=13pt}
\newcommand{\smalllineskip}{\baselineskip=10pt}

\def\eightcirc{
\begin{picture}(0,0)
\put(4.4,1.8){\circle{6.5}}
\end{picture}}
\def\eightcopyright{\eightcirc\kern2.7pt\hbox{\eightrm c}} 

\newcommand{\copyrightheading}[1]
        {\vspace*{-2.5cm}\smalllineskip{\flushleft
        {\footnotesize International Journal of Modern Physics C, #1}\\
        {\footnotesize $\eightcopyright$\, World Scientific Publishing
         Company}\\
         }}

\newcommand{\publisher}[2]{{\begin{center}\footnotesize\smalllineskip 
        Received #1\\
        Revised #2
        \end{center}
        }}

\def\abstracts#1#2#3{{
        \centering{\begin{minipage}{4.5in}\baselineskip=10pt\footnotesize
        \parindent=0pt #1\par 
        \parindent=15pt #2\par
        \parindent=15pt #3
        \end{minipage}}\par}} 

\def\keywords#1{{
        \centering{\begin{minipage}{4.5in}\baselineskip=10pt\footnotesize
        {\footnotesize\it Keywords}\/: #1
        \end{minipage}}\par}}

\newcommand{\bibit}{\nineit}
\newcommand{\bibbf}{\ninebf}
\renewenvironment{thebibliography}[1]
        {\frenchspacing
         \ninerm\baselineskip=11pt
         \begin{list}{\arabic{enumi}.}
        {\usecounter{enumi}\setlength{\parsep}{0pt}     
         \setlength{\leftmargin 12.7pt}{\rightmargin 0pt} 
         \setlength{\itemsep}{0pt} \settowidth
        {\labelwidth}{#1.}\sloppy}}{\end{list}}

\newcounter{itemlistc}
\newcounter{romanlistc}
\newcounter{alphlistc}
\newcounter{arabiclistc}

\newcommand{\fcaption}[1]{
        \refstepcounter{figure}
        \setbox\@tempboxa = \hbox{\footnotesize Fig.~\thefigure. #1}
        \ifdim \wd\@tempboxa > 5in
           {\begin{center}
        \parbox{5in}{\footnotesize\smalllineskip Fig.~\thefigure. #1}
            \end{center}}
        \else
             {\begin{center}
             {\footnotesize Fig.~\thefigure. #1}
              \end{center}}
        \fi}

\newcommand{\tcaption}[1]{
        \refstepcounter{table}
        \setbox\@tempboxa = \hbox{\footnotesize Table~\thetable. #1}
        \ifdim \wd\@tempboxa > 5in
           {\begin{center}
        \parbox{5in}{\footnotesize\smalllineskip Table~\thetable. #1}
            \end{center}}
        \else
             {\begin{center}
             {\footnotesize Table~\thetable. #1}
              \end{center}}
        \fi}

\def\@citex[#1]#2{\if@filesw\immediate\write\@auxout
        {\string\citation{#2}}\fi
\def\@citea{}\@cite{\@for\@citeb:=#2\do
        {\@citea\def\@citea{,}\@ifundefined
        {b@\@citeb}{{\bf ?}\@warning
        {Citation `\@citeb' on page \thepage \space undefined}}
        {\csname b@\@citeb\endcsname}}}{#1}}

\newif\if@cghi
\def\cite{\@cghitrue\@ifnextchar [{\@tempswatrue
        \@citex}{\@tempswafalse\@citex[]}}
\def\citelow{\@cghifalse\@ifnextchar [{\@tempswatrue
        \@citex}{\@tempswafalse\@citex[]}}
\def\@cite#1#2{{$\null^{#1}$\if@tempswa\typeout
        {IJCGA warning: optional citation argument 
        ignored: `#2'} \fi}}

\def\pmb#1{\setbox0=\hbox{#1}
        \kern-.025em\copy0\kern-\wd0
        \kern.05em\copy0\kern-\wd0
        \kern-.025em\raise.0433em\box0}

\def\fnt#1#2{\footnotetext{\kern-.3em
        {$^{\mbox{\scriptsize #1}}$}{#2}}}

\def\fpage#1{\begingroup
\voffset=.3in
\thispagestyle{empty}\begin{table}[b]\centerline{\footnotesize #1}
        \end{table}\endgroup}

\def\runninghead#1#2{\pagestyle{myheadings}
\markboth{{\protect\footnotesize\it{\quad #1}}\hfill}
{\hfill{\protect\footnotesize\it{#2\quad}}}}
\font\tenrm=cmr10
\font\tenit=cmti10 
\font\tenbf=cmbx10
\font\bfit=cmbxti10 at 10pt
\font\ninerm=cmr9
\font\nineit=cmti9
\font\ninebf=cmbx9
\font\eightrm=cmr8

\def\qed{\hbox{${\vcenter{\vbox{                        
   \hrule height 0.4pt\hbox{\vrule width 0.4pt height 6pt
   \kern5pt\vrule width 0.4pt}\hrule height 0.4pt}}}$}}

\renewcommand{\thefootnote}{\fnsymbol{footnote}}        

\def\bsc{{\sc a\kern-6.4pt\sc a\kern-6.4pt\sc a}}       
\def\bflatex{\bf L\kern-.30em\raise.3ex\hbox{\bsc}\kern-.14em 
T\kern-.1667em\lower.7ex\hbox{E}\kern-.125em X} 

\begin{document}
\runninghead{R. Faller, M. P\"utz \& F. M\"uller-Plathe}
{Orientation Correlation In Simplified Models Of Polymer Melts}
\normalsize\textlineskip
\thispagestyle{empty}
\setcounter{page}{1}
\copyrightheading{Vol. 0, No. 0 (0000) 000--000}
\vspace*{0.88truein}
\fpage{1}
\centerline{\bf ORIENTATION CORRELATION IN SIMPLIFIED MODELS}
\vspace*{0.035truein}
\centerline{\bf OF POLYMER MELTS}
\centerline{\footnotesize ROLAND FALLER, MATHIAS P\"UTZ  and FLORIAN
  M\"ULLER-PLATHE} 
\vspace*{0.015truein}
\centerline{\footnotesize\it Max-Planck-Institut f\"ur Polymerforschung,}
\baselineskip=10pt
\centerline{\footnotesize\it Ackermannweg 10, D-55128 Mainz, Germany}
\vspace*{0.15truein}
\vspace*{0.225truein}
\publisher{}{}
\vspace*{0.21truein} 
\abstracts{
  We investigate mutual local chain order in systems of fully flexible
  polymer melts in a simple generic bead-spring model. The excluded-volume
  interaction together with the connectivity leads to local ordering effects
  which are independent of chain length between 25 and 700 monomers, i.e. in
  the Rouse as well as in the reptation regime. These ordering phenomena
  extend to a distance of about 3 to 4 monomer sizes and decay to zero
  afterwards.}{}{} 

\vspace*{1pt}\textlineskip
\section{Introduction}
\noindent
Nuclear magnetic resonance (NMR) experiments on polymer melts and glasses show
effects of mutual local ordering
of neighboring chains \cite{graf98,havens83}. These effects are found far
away from the glass transition in non liquid-crystalline polymer melts like
polybutadiene in the reptation regime. Ordering effects in polymer melts were
already investigated in several molecular dynamics simulations of
specific polymer melts \cite{smith94a,smith94b,moe95b}, block copolymers
\cite{trohalaki96} and by Monte Carlo methods using lattice models
\cite{kolinski86}. However, to date,  only very few
investigations of local order in small systems were performed\cite{rigby88} in
generic off-lattice models, 
although these models are rather well investigated in other contexts
\cite{kremer89}. It is not clear how much influence depends on the specific
polymer interactions  and how much orientational
correlation is contributed by the excluded volume interaction which may be
modeled 
regardless of the specific chemical system. In this work, we focus on the
generic effects of local order in fully flexible polymer melts.  
\eject
\textheight=7.7truein
\setcounter{footnote}{0}

\renewcommand{\thefootnote}{\alph{footnote}}
\section{Method}
\noindent
We performed molecular dynamics simulation of fully flexible  polymer chains
using a truncated and shifted Lennard-Jones potential
(Weeks-Chandler-Anderson potential) for the excluded volume interaction
between all beads
\begin{equation}
V_{LJ}(r)=4\epsilon\left[\left(\frac{\sigma}{r}\right)^{12}-
\left(\frac{\sigma}{r}\right)^{6}\right], r=r_{c}=\sqrt[6]{2}\sigma
\end{equation}
and a finitely extendable non-linear elastic (FENE) potential 
\begin{equation}
V_{FENE}(r)=\frac{\alpha}{2}\frac{R^{2}}{\sigma^{2}}
\ln\left(1-\frac{r^{2}}{R^{2}}
\right), r<R=1.5\sigma, \alpha =30\epsilon
\end{equation}
to model the connectivity along the chains. The system uses orthorombic
periodic boundary conditions. Temperature is kept constant by 
simulation of a Langevin equation with a stochastic force and a friction term
added to Newton's equation of motion. The simulations were performed at a
number density of $\rho=0.85$ in reduced units and a temperature of
$T=1.0$ with a friction strength of $\xi=0.5$. The program used is described
in detail in reference\cite{puetz98}. Our melt simulations consist of chains
with 25 to 700 monomers. The overall number of monomers in the simulation
lies between 80.000 and 500.000. 

We investigated the correlation between unit vectors defined along
the chain 
\begin{equation}
\vec{u}_{d}=\vec{r_{i}}-\vec{r}_{i-d}.
\end{equation}
A value of $d=1$ means therefore just bond vectors between connected
beads. The static correlation of these unit vectors is calculated. The 2nd
Legendre polynomial of the scalar product 
\begin{equation}
P_{2}(r)=\frac{1}{2}\langle 3(\vec{u}(0)\cdot\vec{u}(r))^{2}-1\rangle
\end{equation}
was recorded versus
the distance of the center of mass of the two chain segments and averaged over
all corresponding pairs with equal distance. We chose this orientational
distribution function (ODF) because it is the relevant
quantity in NMR measurements and it makes no sense to distinguish between the
two possible directions of the  chains. The orientational distribution
function is 
negative when the two vectors are perpendicular and
positive for parallel orientation. If the vectors are uncorrelated the
function is zero. In our investigations, we only looked at the inter-chain
effects. All correlations along one chain were explicitly excluded.

\section{Results and Discussion}
\noindent
The orientational distribution functions of chains at length
25 and 700 are shown in figure \ref{fig:ocf} on the left hand side for
$d=1$. The entanglement length of this system was earlier found to be 35 beads
\cite{kremer89} whereas recent simulations are more consistent with 60.
The simulation with 25 and 50 monomers is in the Rouse regime, simulations
with 100 or 200 monomers are in the region of crossover, whereas
the longest simulations (700 monomers) are  in the reptation regime. The
structure and the 
numerical values of the peaks are almost indistinguishable between different
chain lengths. This shows that the ordering phenomenon is strictly local and is
probably imposed by chain packing. The initial negative region indicates
nearly perfect perpendicular orientation. This may be understood if one
realizes that the bond center distance is clearly smaller than the distance of
the first peak
in the radial distribution function ($\sqrt[6]{2}$). Hence only a perpendicular
alignment is compatible with this small distance. Such close contacts,
however, are
extremely rare. The radial distribution function here is almost zero.
The first positive peak in the ODF coincides rather well with the first peak
in the interchain radial distribution function (figure \ref{fig:ocf} right
hand side, solid line). The depression of $g(r)_{inter}$ at low distances is
due to the enhanced density of the same chain in the vicinity of a monomer
(correlation hole).

At the distance of the first 
neighbor shell, the monomers are preferentially parallel oriented. This is
consistent with results on short chains of only 10 or 20 monomers\cite{rigby88}.
The values of $P_{2}$ of individual pairs of vectors vary in a wide range
between strongly negative and strongly positive values. Only a slight
preference of the parallel orientation leads to this peak. This preferred
alignment is, in general, not mediated by an intermediate perpendicular chain.

\begin{minipage}[b]{7.02cm}
\parbox[h]{7.01cm}{
\begin{center}
\psfig{figure=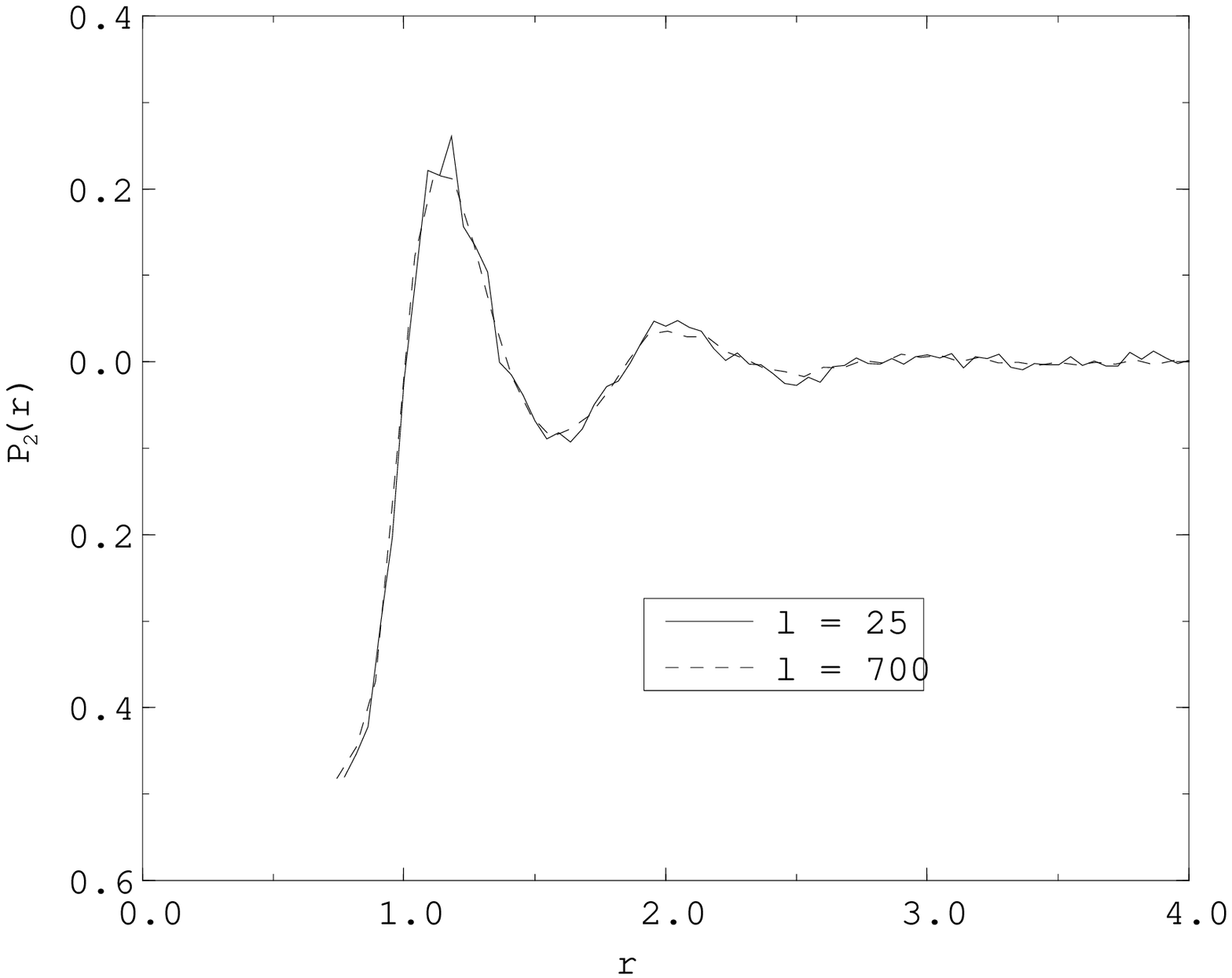,width=5cm} 
\end{center}
}
\end{minipage}\hfill
\begin{minipage}[b]{7.02cm}
\parbox[h]{7.01cm}{
\begin{center}
\psfig{figure=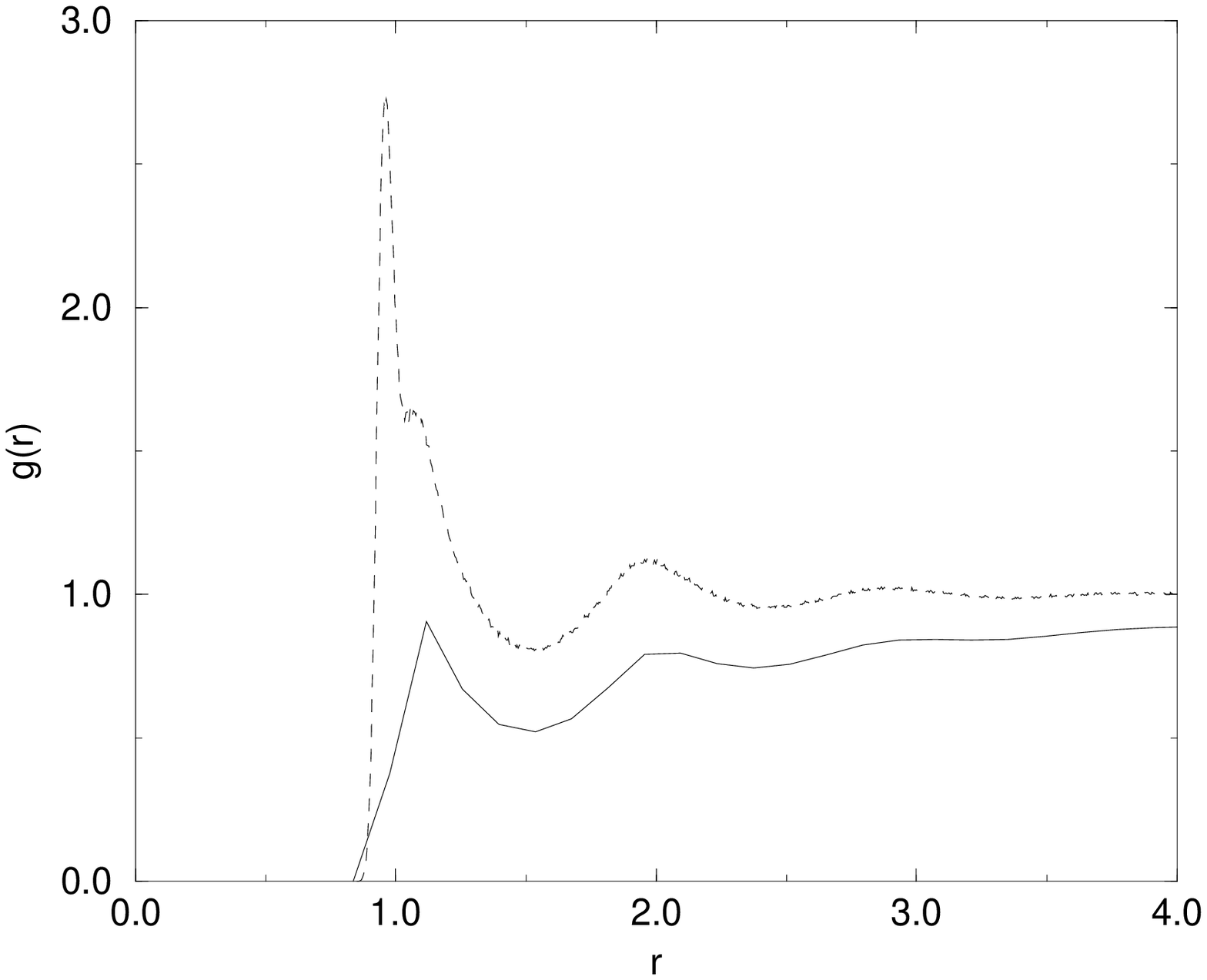,width=5.5cm}
\end{center}
}
\end{minipage}

\vspace*{10pt}
\fcaption{left: Orientational distribution functions of chains with 25 (3200
  chains), 50 (5000 chains), 200 (2500 chains ) and 700 (500 chains) monomers
  for vectors connecting next neighbors.\\
  right: Radial distribution functions $g(r)$ of a melt with 2500 chains of 200
  monomers. Dashed line: overall $g(r)$, solid line: interchain $g(r)$.}
\label{fig:ocf}
\vspace*{13pt}

\noindent
Orientational correlations are visible up to a distance of about 3 monomer
radii. The ODF falls to zero at longer distances, showing that there is no
nematic long-range ordering which of course is not expected in fully
flexible systems. Figure \ref{fig:localpict} shows typical local arrangements
which may lead to the observed orientational distribution functions.

\begin{minipage}[b]{7.02cm}
\parbox[h]{7.01cm}{
\begin{center}
\psfig{file=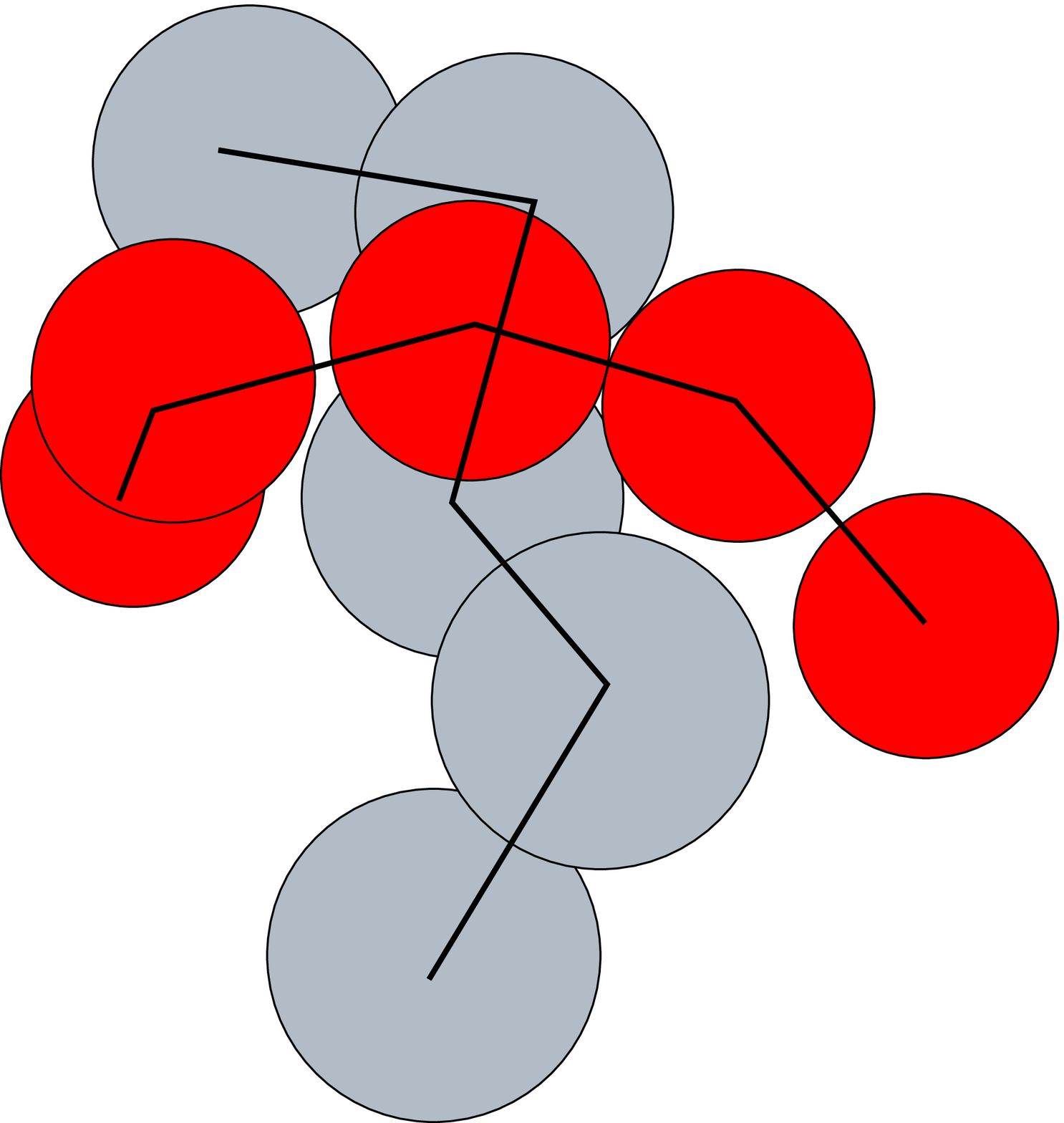,width=5cm}
\end{center}
}
\end{minipage}\hfill
\begin{minipage}[b]{7.02cm}
\parbox[h]{7.01cm}{
\begin{center}
\psfig{file=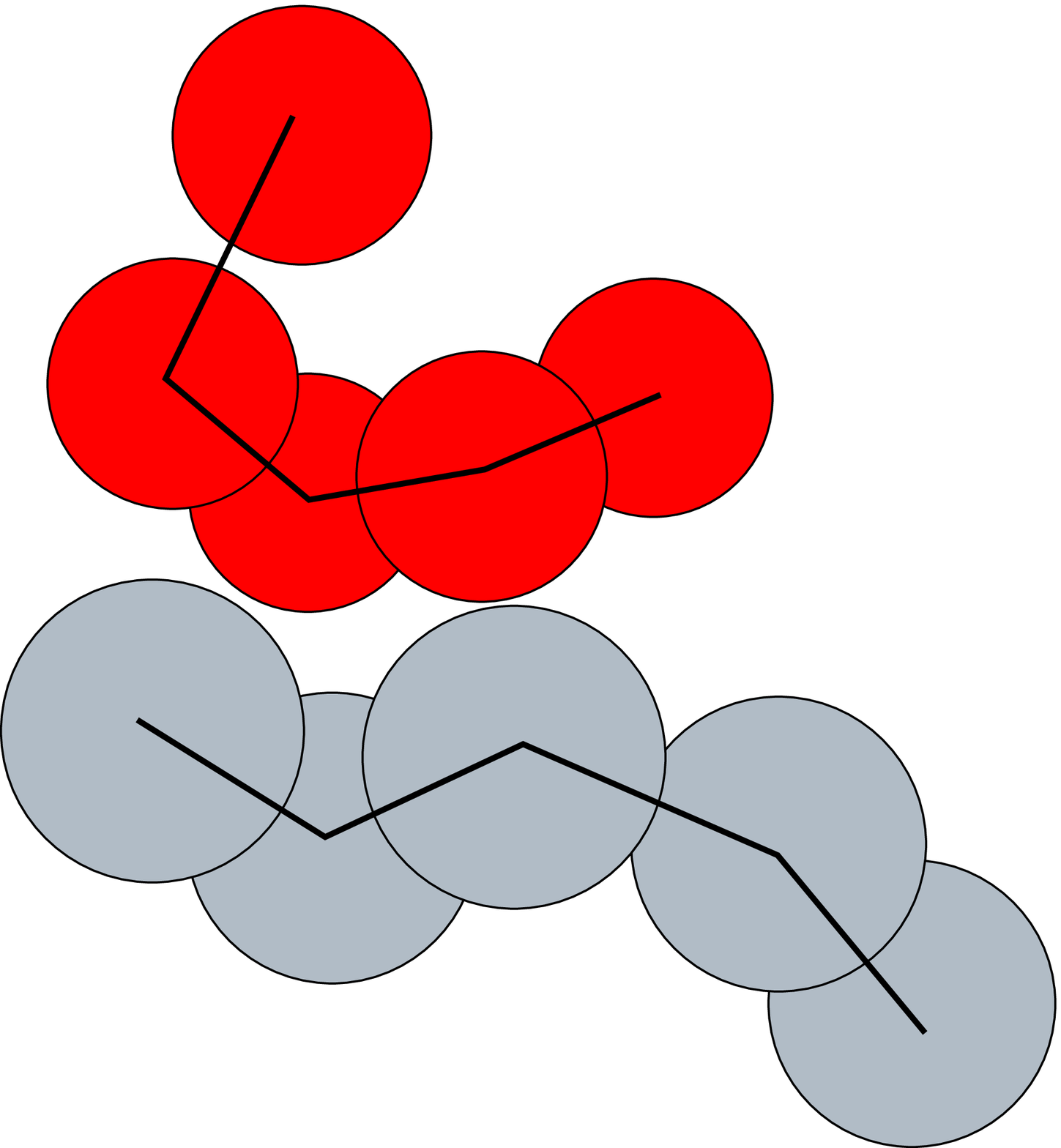,width=5cm}
\end{center}
}
\end{minipage}

\vspace*{10pt}
\fcaption{Selected arrangements leading to negative (left) or positive (right)
  orientational distribution functions.}
\label{fig:localpict}
\vspace*{13pt}

\noindent
We also investigated the ordering of longer chain segments. We
defined unit vectors in the direction from one monomer to its next-nearest
neighbor $(d=2)$, next to next-nearest $(d=3)$ etc. The orientational
correlations are 
weaker but remain visible (see figure \ref{fig:different-d}).

\begin{minipage}[b]{7.02cm}
\parbox[h]{7.01cm}{
\begin{center}
\psfig{file=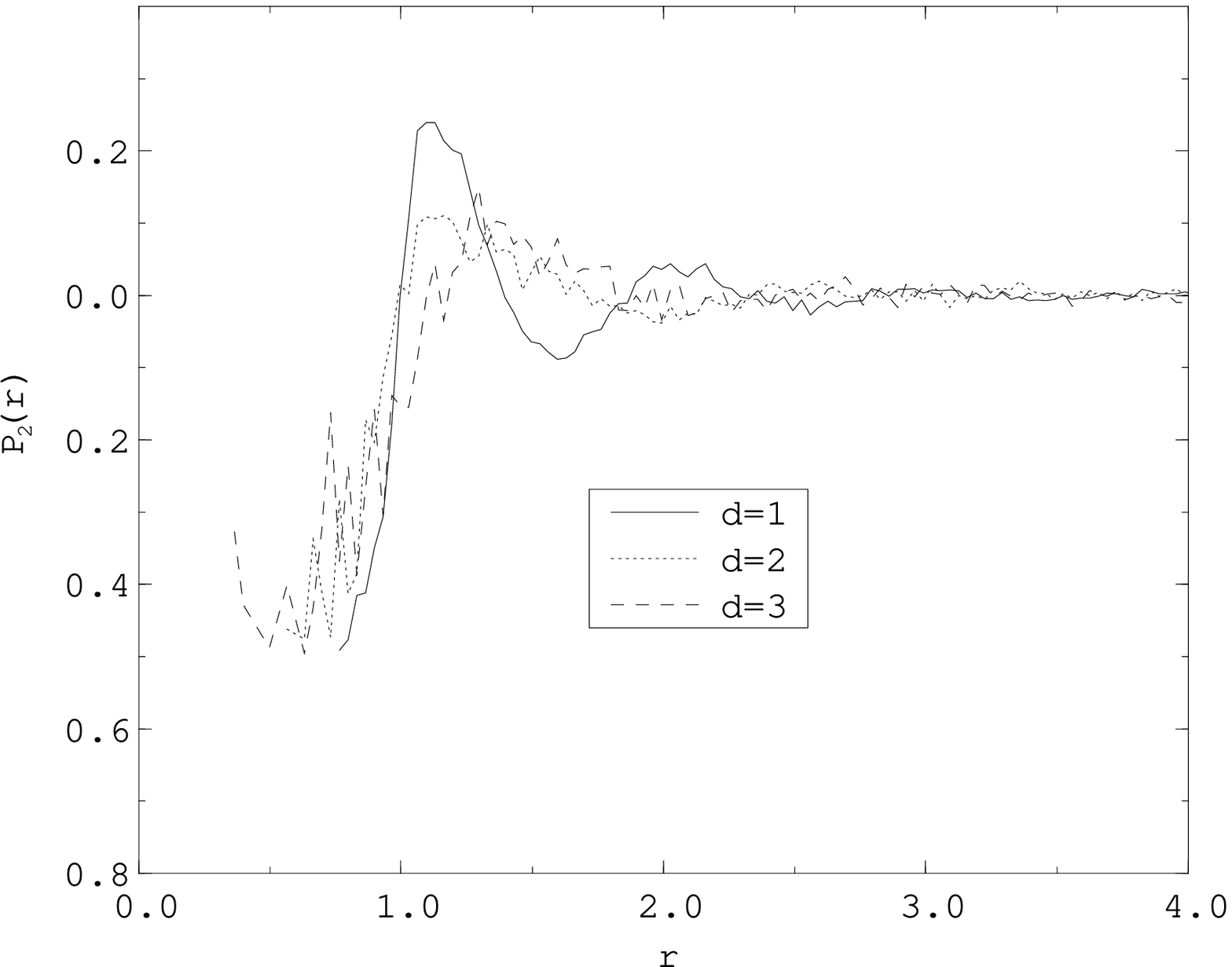,width=5cm}
\end{center}
}
\end{minipage}\hfill
\begin{minipage}[b]{7.02cm}
\parbox[h]{7.01cm}{
\begin{center}
\psfig{file=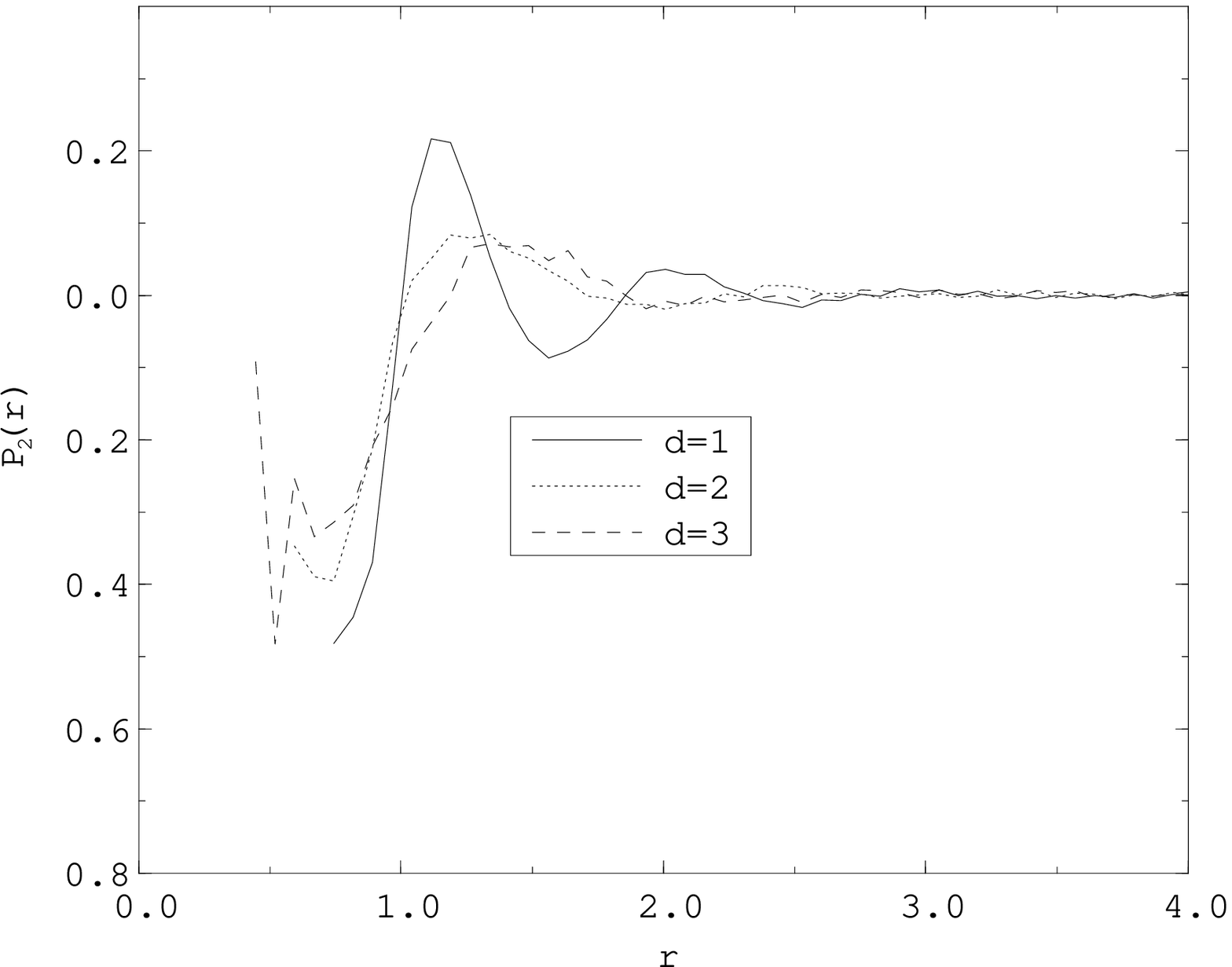,width=5cm}
\end{center}
}
\end{minipage}

\vspace*{10pt}
\fcaption{Orientational distribution functions of chains with 50 monomers (5000
  chains,left) and 700 monomers (500 chains, right) depending on segment
  length $d$ (see text)}
\label{fig:different-d}
\vspace*{13pt}

\noindent
The first peak is shifted to longer distances which is not surprising because
of the bigger segment size. This shows that the local ordering of the bonds
also leads to an ordering of segments of several beads. The maximum of the ODF
for $l=700$ and $d=2,3$ seems to be at a slightly bigger distance, but this is
not yet significant. The orientational distribution functions are still too
noisy to decide this clearly.

To summarize, we find local ordering of adjacent polymer chains in the melt
even for a 
fully flexible model with no interactions other than excluded volume.
This ordering does not depend on the overall chain length since it is a local
property. Even
the otherwise drastic difference between entangled and unentangled systems does
not influence the static correlations. Presumably, the lifetime of the
correlations is shorter than the time-scale relevant for reptation.

\nonumsection{Acknowledgments}
\noindent
Fruitful discussions with A. Heuer and A. Kolb are gratefully acknowledged.

\nonumsection{References}

\end{document}